\documentclass[aps,preprint]{revtex4}
\usepackage{graphicx}
\begin{document}
\title {\Large \bf  Electromagnetic Properties of Kerr-anti-de Sitter Black Holes}
\author{\large Alikram N. Aliev}
\address{Feza G\"ursey Institute, P. K. 6  \c Cengelk\" oy, 34684 Istanbul, Turkey}
\date{\today}

\begin{abstract}

We  examine  the electromagnetic properties of  Kerr-anti-de Sitter (Kerr-AdS) black holes in four and higher spacetime dimensions. Assuming that the black holes may carry a test electric charge we show that the Killing one-form which represents the {\it difference} between the timelike generators in the spacetime and in the reference
background can be used as a potential one-form for the associated
electromagnetic field. In four dimensions the potential one-form
and the Kerr-AdS metric with properly re-scaled mass parameter
solve the Einstein-Maxwell equations, thereby resulting in the
familiar Kerr-Newman-AdS solution. We solve the quartic equation
governing the location of the event horizons of the
Kerr-Newman-AdS  black holes and present closed analytic
expressions for the radii of the horizons. We also compute the
gyromagnetic ratio for these black holes and show that it
corresponds to $\,g=2\,$ just as for ordinary black holes in
asymptotically flat spacetime. Next, we compute the gyromagnetic
ratio for the Kerr-AdS  black holes with a single angular momentum
and with a test electric charge in all higher dimensions. The
gyromagnetic ratio crucially depends on the dimensionless ratio of
the rotation parameter to the curvature radius of the AdS
background. At the critical limit, when the boundary Einstein
universe is rotating at the speed of light, it tends to $ g=2 $
irrespective  of the spacetime dimension. Finally, we consider the
case of a five dimensional Kerr-AdS black hole with two angular
momenta and show that it possesses two distinct gyromagnetic
ratios in accordance with its two orthogonal 2-planes of rotation.
In the special case of two equal angular momenta, the two
gyromagnetic ratios merge into one leading to  $ g=4 $ at the
maximum angular velocities of rotation.

\end{abstract}
\maketitle

\section{Introduction}

Black holes remain  one of the most important objects of study in
four and higher dimensions. We begin with higher-dimensional black
holes in asymptotically flat spacetimes. Of particular interest is
the role of the black holes in string/M-theory, since they may
serve as theoretical laboratories to test the novel predictions of
the theory.  The intriguing example of this is the
statistical-mechanical explanation of the Bekenstein-Hawking
entropy for certain supersymmetric black holes in five dimensions
\cite{bmpv}. Developments have revealed new unexpected features of
black holes in higher dimensions. It turns out that higher
dimensions endow black holes with different horizon topologies,
whereas in four dimensions the horizon is uniquely determined by
the topology of a two-sphere \cite{Israel}. Some basic properties
of black holes in four dimensions such as  their {\it stability}
and {\it uniqueness} properties also change in higher dimensions.

The simplest metrics for black holes with spherical horizon
topology in all higher dimensions were found by Tangherlini many
years ago \cite{tang}. These metrics generalize the familiar
nonrotating Schwarzschild and Reissner-Nordstr\"om solutions of
four-dimensional general relativity. The rotational dynamics of
these black holes was first explored by Myers and Perry, who
discovered the general rotating black hole solutions in arbitrary
dimensions \cite{mp}. The Myers-Perry solution is not unique,
unlike its four-dimensional counterpart, the Kerr solution. There
exists a rotating black ring solution with the horizon topology of
$\, S^2\times S^1\,$ which may have the same mass and spin as the
Myers-Perry solution in five dimensions \cite{er1} (see also Ref.
\cite{er2}). In this sense, the five-dimensional spacetime harbors
a relative of the Myers-Perry black hole in the form of a ``
donut-shaped " rotating black hole. The different properties of
black holes and black rings as well as new exact solutions have
been discussed in \cite{fs1}-\cite{yazad}.

Higher-dimensional black holes have a rich phenomenology. This
became fully transparent after the advent of Large Extra Dimension
Scenarios \cite{ADD}. These scenarios are built on the idea that
our observable universe is a slice, a {\it braneworld }, in a
higher-dimensional spacetime. The braneworld black holes are in
general higher-dimensional objects as they must carry the imprint
of the extra dimensions. \cite{chr}-\cite{aemir}. The Large Extra
Dimension Scenarios open up the possibility of directly probing
TeV-scale mini black holes in high-energy collisions \cite{gt}.
For other related works on classical and quantum properties of
black holes in braneworld scenarios, see Refs.
\cite{dejan}-\cite{majumdar}.

Developments in string/M-theory have also greatly stimulated the
study of black hole solutions in anti-de Sitter space. The
striking examples of this come from the AdS/CFT correspondence
between a weakly coupled gravity system in an AdS background and a
strongly coupled conformal field theory (CFT) on its boundary
\cite{mal}. It is well known that the familiar Schwarzschild
solution in AdS space describes a simple nonrotating AdS black
hole. The most important feature of the black hole is that it has
a minimum critical temperature determined by the curvature radius
of the AdS background. This means that there must be a thermal
phase transition between AdS space and Schwarzschild-AdS space at
a fixed temperature: At low temperatures thermal radiation in the
AdS space is in stable equilibrium, while at temperatures higher
than the critical value there is no stable equilibrium
configuration without a black hole \cite{hpage}. The Hawking-Page
transition was interpreted by Witten \cite{witten} in terms of a
transition between the confining and deconfining phases of the
corresponding conformal field theory. As an interesting
application of the AdS/CFT correspondence, this result has been
extensively discussed in the context of both static and rotating
AdS black holes in various spacetime dimensions (see Refs.
\cite{emparan2}-\cite{odin}). It is also worth to note stability
properties of AdS black holes in four dimensions. It has been
shown that large Kerr-AdS  black holes are always classically
stable, while small Kerr-AdS  black holes become unstable with
respect to scalar and gravitational perturbations via the
superradiant amplification mechanism \cite{hreal, cardi}.

The authors of Ref. \cite{hhtr} have studied the relationship
between Kerr-AdS black holes in the bulk and conformal field
theory living on a boundary Einstein universe. Probing  the
AdS/CFT correspondence in the critical limit, in which the
rotation of the boundary Einstein universe occurs at the speed of
light, they found that the generic thermodynamic features of the
conformal field theory agree with those of the black holes in the
bulk. In general, the clearest description of the boundary
conformal field theory in a rotating Einstein universe imprinted
by a Kerr-AdS black hole in the bulk is a very complicated and
subtle question. Exploring the critical limit where the boundary
Einstein universe rotates at the speed of light makes a
significant simplification as it incorporates generic features of
both bulk and boundary theories. There are subtleties even with
the definition of the total mass and angular velocities of the
Kerr-AdS black holes. In Ref. \cite{gpp1},  it has been argued
that one must evaluate the mass and angular velocities relative to
a frame which is nonrotating at infinity. Only these quantities
define the most important characteristics of the Kerr-AdS black
holes relevant for their CFT duals and satisfy the first law of
thermodynamics. A more detailed analysis has led to a general
equivalence between the bulk and the boundary thermodynamic
variables \cite{gpp2,gpp3}, thereby clarifying the lack of
unanimity in many previous cases \cite{bala1}.

In the light of these developments, we address further important
properties of the black holes in AdS space, namely the
electromagnetic properties of the Kerr-AdS black holes carrying a
test electric charge. In order to construct the associated
solution of the Maxwell field equations, we use the {\it
difference} between the timelike generators in the Kerr-AdS metric
and in the reference background. The latter is taken to be a
rotating Einstein universe. We also compute the magnetic dipole
moments and the gyromagnetic ratios for the rotating charged AdS
black holes. We consider the following cases: Kerr-Newman-AdS
black holes in four dimensions, Kerr-AdS black holes with a single
angular momentum in all higher dimensions and Kerr-AdS black holes
with two angular momenta in five dimensions. The basic result for
the gyromagnetic ratio of the Kerr-AdS black holes with a single
angular momentum is given in \cite{aliev3}. Here we present full
details of this result.

In classical electrodynamics the gyromagnetic ratio $ g $  relates
the magnetic dipole moment of a  charged rotating body to its
total angular momentum and  $ g=1 $ for a constant ratio of the
charge to mass density. However, quantum electrodynamics predicts,
up to radiative corrections, that $\,g=2\,$ for charged fermions,
like electrons and muons. It has been shown that at the
tree-level, $\,g=2\,$ is the natural value of the gyromagnetic
ratio for elementary particles of arbitrary spin \cite{fmt}. The
exact result $\,g=2\,$ is related to unbroken supersymmetry and
the factor $\,g-2\,$ is considered to be a measure of supersymmetry-breaking
(SUSY-breaking) effects (see Ref. \cite{dom} for a recent review).
From a classical point of view, it is also clear that a rotating
charged black hole must have a magnetic dipole moment. However,
the magnetic dipole moment is not an independent quantity but is
determined by the mass, angular momentum and the electric charge
of the black hole. It is long known that, unlike a uniformly
charged rotating body, the gyromagnetic ratio for  a rotating and
charged black hole in general relativity is equal to $2$, the same
value as for an electron in the Dirac theory \cite{carter1}. In
further developments, this remarkable fact has been confirmed in
many cases of Einstein-Maxwell fields in four dimensions
\cite{rt,bahram}.

In recent works \cite{af,aliev1}, the gyromagnetic ratio was
studied in higher dimensions for asymptotically flat Myers-Perry
black holes carrying a test electric charge as well as for
arbitrary values of the electric charge in the limit of slow
rotation. A detailed numerical treatment of the problem  was given
in Refs. \cite{kunz}.  It should be noted that, unlike four
dimensions, the value of the gyromagnetic ratio is not universal
in higher dimensions. For a five dimensional Myers-Perry black
hole with a test electric charge the gyromagnetic ratio was found
to be $ g=3 $. Earlier, the same value of the gyromagnetic ratio
was found for a supersymmetric rotating black hole in five
dimensions \cite{herdeiro}. The gyromagnetic ratio of black rings
was studied in \cite{opravda}. On the other hand, it is known that
for black holes in five-dimensional Kaluza-Klein theory the
$g$-factor approaches unity in the ultrarelativistic limit
\cite{gw}. This value is the natural $g$-factor for massive states
in the Kaluza-Klein theory \cite{hosoya}. It is also known that
for Kaluza-Klein black holes in ten-dimensional supergravity $ g=1
$ \cite{duff}, while some $ p\,$-brane solutions in higher
dimensions have a gyromagnetic ratio that corresponds to $ g=2 $
\cite{bala2}.

The  present paper is organized as follows. In Sec. II we begin
with a brief description of the basic properties of the Kerr-AdS
metric in four dimensions. We give the definition of the angular
velocity for {\it locally nonrotating} observers in this spacetime
and obtain the expressions for the mass, angular momentum and
angular velocity that are consistent with the first law of
thermodynamics. We put a test electric charge on the black hole
and construct the corresponding solution of the Maxwell equations
in the Kerr-AdS background. In Sec. III we assume that the black
hole may have an arbitrary electric charge.  In this case,  by an
appropriate re-scaling of the mass parameter one can pass from the
Kerr-AdS solution to the Kerr-Newman-AdS one in which the
electromagnetic field is still given by the expressions found in
Sec. II within the `` test-charge " approach. Next, we solve the
quartic equation governing the location of the event horizons of
the Kerr-Newman-AdS  black holes and present closed analytic
expressions for the radii of the horizons. We also compute the
gyromagnetic ratio for these black holes and show that it
corresponds to the value $ g=2 $ irrespective of the AdS nature of
the spacetime. The electromagnetic properties of rotating Kerr-AdS
black holes with a single angular momentum in all higher
dimensions are studied in Sec. IV. Here we extend the test-charge
approach to higher dimensions and find the potential one-form  for
the electromagnetic field generated by a test electric charge of
the black holes. We also compute the magnetic dipole moment and
the gyromagnetic ratio for these black holes using
thermodynamically consistent expressions for the mass and angular
momentum. The value of the gyromagnetic ratio crucially depends on
the dimensionless ratio of the rotation parameter to the curvature
radius of the AdS spacetime. In the critical limit in which the
boundary Einstein universe is rotating at the speed of light, the
gyromagnetic ratio approaches  $ g=2 $ regardless of the spacetime
dimension. It is known that at the critical limit of rotation, the
Kerr-AdS black holes are related to SUSY configurations
\cite{cvetic}. Thus, it follows from our result that a
supersymmetric black hole in an AdS background must have the
gyromagnetic ratio corresponding to $g=2$. Finally, in Sec. V we
consider a general five-dimensional Kerr-AdS black hole with two
independent angular momenta and with a test electric charge. We
obtain the precise expressions for the angular velocities of
locally nonrotating  observers and re-derive the expressions for
the mass and angular momenta. We construct the potential one-qform
that describes the test electromagnetic field of the black holes.
Here we also define a natural orthonormal frame in which the
electromagnetic field two-form takes its simplest form. This is a
generalization of the familiar Carter frame in four-dimensional
Kerr-Newman spacetime. We show that a five dimensional charged
Kerr-AdS black hole possesses two distinct gyromagnetic ratios. In
the special case of two equal angular momenta, the two
gyromagnetic ratios merge into one which tends to
 $ g=4 $ for the maximally rotating boundary of the spacetime.

\section{Kerr-AdS black holes in four dimensions }

The exact solution of the Einstein field equations with a
cosmological constant that describes rotating  black holes in
four-dimensional spacetime with asymptotic (anti)-de Sitter
behavior was found in \cite{carter2}. The corresponding spacetime
metric in the Boyer-Lindquist coordinates has the form
\begin{eqnarray}
ds^2 & = & -{{\Delta_r}\over {\Sigma}} \left(\,dt - \frac{a
\sin^2\theta}{\Xi}\,d\phi\,\right)^2 + {\Sigma \over~ \Delta_r}
dr^2 + {\Sigma \over ~\Delta_{\theta}}\,d\theta^{\,2} +
\frac{\Delta_{\theta}\sin^2\theta}{\Sigma} \left(a\, dt -
\frac{r^2+a^2}{\Xi} \,d\phi \right)^2\,\,, \label{4kads}\nonumber\\
& &
\end{eqnarray}
where
\begin{eqnarray}
\Delta_r &= &\left(r^2 + a^2\right)\left(1 +\frac{r^2}{l^2}\right)
- 2 M r \,,~~~~~~~\Sigma = r^2+ a^2 \,\cos^2\theta \,,
\nonumber \\[2mm]
\Delta_\theta & = & 1 -\frac{a^2}{l^2}
\,\cos^2\theta\,,~~~~~~~~~~~~~~~~~~~~~~~~~ \Xi=1 -
\frac{a^2}{l^2}\,\,. \label{metfunct}
\end{eqnarray}
The parameters $M$ and $a$ are related to the mass and angular
momentum of the black hole,  $l$ is the curvature radius
determined by the negative cosmological constant $\, \Lambda= -3
l^{-2}\,$.  The metric determinant is given by
\begin{equation}
\sqrt{-g}=\frac{\Sigma \sin\theta}{\Xi}\,. \label{4ddet}
\end{equation}
Clearly, the rotation parameter $ a $ must satisfy the relation
$\,a^2 < l^2\,,$ but when approaching the critical limit $\,a^2 =
l^2\,$, the metric becomes singular. In this limit the boundary of
AdS spacetime, which is a three-dimensional Einstein universe,
rotates at the speed of light.

The stationarity and rotational symmetry properties of the metric
(\ref{4kads}) imply the existence of two commuting Killing vector
fields
\begin{equation}
\xi_{(t)}=\frac{\partial}{\partial t}\,\,, ~~~~~~~~~\xi_{(\phi)}=
\frac{\partial}{\partial \phi}\,\,. \label{killing}
\end{equation}
The various scalar products of these Killing vectors can be
expressed through the metric components as follows
\begin{eqnarray}
{\bf \xi}_{(t)} \cdot {\bf \xi}_{(t)}&=& g_{tt}= -1+
\frac{2 M r}{\Sigma}- \frac{r^2+a^2 \sin^2\theta}{l^2}\,\,,\nonumber \\[3mm]
{\bf \xi}_{(t)} \cdot {\bf \xi}_{(\phi)}&= &g_{t\phi}= \,\frac{a
\sin^2 \theta}{\Xi}\left(\frac{r^2+a^2}{l^2}-  \frac{2 M
r}{\Sigma}\right)\,\,,
\\[3mm]
{\bf \xi}_{(\phi)} \cdot {\bf \xi}_{(\phi)}&=& g_{\phi\phi}=
\,\frac{\sin^2 \theta}{\Xi^2}\, \left[ \left(r^2+a^2\right)\Xi +
\frac{2 M r a^2 \sin^2\theta}{\Sigma}\right] \nonumber\,\,.
\label{kproduct}
\end{eqnarray}
Another important feature of the Kerr-AdS spacetime  becomes
transparent when  one introduces  a family of  locally nonrotating
observers that move on  orbits with constant $r$ and $\theta$ and
with a four-velocity $ u^{\mu} $ such that $ u\cdot
{\xi}_{(\phi)}=0 $. The  coordinate angular velocity of these
observers is given by
\begin{eqnarray}
\Omega&=& -\frac{g_{t \phi}}{g_{\phi \phi}}= \frac{a \,\Xi\,
\left[\left(r^2+a^2\right) \Delta_{\theta}-
\Delta_{r}\right]}{\Gamma}\,\,, \label{angvelocity1}
\end{eqnarray}
where
\begin{equation}
 \Gamma= \left(r^2+a^2\right)^2 \,\Delta_{\theta} - \Delta_{r}
a^2 \sin^2\theta\,\,.
\end{equation}
It is easy to see that, in contrast to the case of an ordinary
Kerr black hole in asymptotically  flat spacetime, the angular
velocity does not vanish at spatial infinity. Instead, we have the
expression
\begin{equation}
\Omega_{\infty}= -\frac{a}{l^2}\,\,. \label{asymangvel}
\end{equation}
This means that the Kerr-AdS metric (\ref{4kads}) is given in a
coordinate system which is rotating at spatial infinity. When
approaching the horizon of the black hole, $\,r \rightarrow
r_{+}\,(\,\Delta_r=0 \,)\,, $ the  angular velocity in
(\ref{angvelocity1}) tends to its constant value
\begin{equation}
\Omega_{H}= \frac{a\,\Xi}{r_{+}^2 + a^2}\,\, \label{hvelocity}
\end{equation}
which can be  thought of as the angular velocity of the black
hole. This is confirmed by the fact that the co-rotating Killing
vector field
\begin{equation}
\chi = \xi_{(t)}+ \Omega_{H}\,\xi_{(\phi)}
\end{equation}
becomes  null at the surface $\,\Delta_r=0 \,$, i.e.  it is
tangent to the null surface of the horizon. Clearly,  one can also
define the angular velocity of the black hole with respect to a
frame that is static at infinity. We have
\begin{eqnarray}
\omega_{H}&=&\Omega_{H}-\Omega_{\infty}=\frac{a}{r_{+}^2 +
a^2}\,\left(1+\frac{r_{+}^2}{l^2}\right)\,\,.
 \label{einvelocity}
\end{eqnarray}
It turns out that this angular velocity is the most important
characteristic of the rotating AdS black holes in the sense that
it enters their consistent thermodynamics \cite{cck}. On the other
hand, it is easy to show that this angular velocity coincides with
that of the boundary Einstein universe \cite{hhtr}, thereby
providing the relevant basis for a CFT dual of the bulk Kerr-AdS
black hole.

Next, we calculate the mass and angular momentum of the metric
(\ref{4kads}). As is known, for black holes in asymptotically flat
spacetime these quantities are unambiguously determined using the
Komar approach \cite{komar}. However, the Komar approach must be
used with care for rotating AdS black holes, since the integral
for the mass gives a divergent result. Therefore, in order to find
a physically meaningful result for the mass one needs to perform a
``background subtraction" that may require  care as well. With all
this in mind, we have the Komar integrals
\begin{eqnarray}
\mathcal{M}^{\prime}&=& - \frac{1}{8 \pi\,}\oint
\,^{\star}d(\delta \hat \xi_{(t)})\,\,,~~~~~~~~~~~J^{\prime}=
\frac{1}{16 \pi\,}\oint \,^{\star}d(\delta \hat
\xi_{(\phi)})\,\,,\label{komar}
\end{eqnarray}
where the $\,{\star}\,$ operator denotes the Hodge dual and the
Killing one-form $\,\hat \xi=\xi_{\mu}\, d x^{\mu}$ is associated
with the Killing vectors in (\ref{killing}). In performing the
above integrals one must integrate the differences $\,\delta \hat
\xi \,$ between the Killing isometries of the spacetime under
consideration and its reference background. For the reference
background, we use the solution (\ref{4kads}) with vanishing mass
parameter ($ M=0 $). We note that this procedure does not
contribute to the result for the angular momentum as the
calculation of the angular momentum is unambiguous in the Komar
method.

Using  the asymptotic expansions of the integrands in equation
(\ref{komar})
\newpage
\begin{eqnarray}
\delta \xi_{(t)}^{t\,;\,r}& = & \frac{M}{r^{2}} +
\mathcal{O}\left(\frac{1}{r^{4}}\right)\,\,,
\nonumber \\[4mm]
\delta \xi_{(\phi)}^{t\,;\,r}& = &- \,\frac{3 a M \sin^2\theta}{
r^2\,\Xi}+ \mathcal{O}\left(\frac{1}{r^{4}}\right)\,\,,
\label{4expansion}
\end{eqnarray}
we perform the integration over a  $2$-sphere at $\,r \rightarrow
\infty\,$. This  gives
\begin{eqnarray}
\mathcal{M}^{\prime}&=& \frac{M}{\Xi}\,\,,~~~~~~ J^{\prime}=
\frac{a M}{\Xi^2}\,\,,\label{4mj}
\end{eqnarray}
where $ J^{\prime} $ is the {\it actual} angular momentum of the
Kerr-AdS  metric. It agrees with the angular momentum obtained
earlier in the literature using different approaches \cite{ht,
perry}. However, the  expression for $ \mathcal{M}^{\prime} $ can
not be regarded as the {\it actual} mass, since it does not
satisfy the first law of thermodynamics. This fact was first
emphasized in \cite{gpp1,gpp2}. The reason for this is the salient
feature of the Kerr-AdS metric (\ref{4kads}) that leads to a
nonvanishing drag at spatial infinity, see Eq. (\ref{asymangvel}).
This fact also means that the timelike Killing vector in
(\ref{killing}), that is used in the calculation of the mass, is
indeed  rotating at infinity. Therefore one must calculate the
mass with respect to a new timelike Killing vector which is
nonrotating at infinity. This Killing vector is
\begin{equation}
\partial_t -\frac{a}{l^2}\,\,\partial_{\phi} \,\,.\label{newkill}
\end{equation}
It is easy to show that it indeed  has the vanishing scalar twist
for $\,M=0\,$. The calculation of the mass becomes transparent
when employing, for instance, the superpotential technique of
Katz, Bi\v c\' ak and Lynden-Bell \cite{kbl}. Adapting it to our
case we have the integral
\begin{equation}
K=- \frac{1}{16 \pi}\oint\, ^{\star}d(\delta \hat \xi) - \frac{1}{8
\pi} \oint \,^{\star}d(\delta S) \,\,, \label{kblpot}
\end{equation}
where
\begin{eqnarray}
S&= &\frac{1}{2} \,\xi_{[\mu} k_{\nu]}\, d x^{\mu}\wedge
dx^{\nu}\,\,,~~~~~~~~~k^{\mu}= g^{\mu \nu }\, \delta
\Gamma^{\lambda}_{\nu \lambda}- g^{\alpha \beta }\, \delta
\Gamma^{\mu}_{\alpha \beta }\,\,.
\end{eqnarray}
We note that $\,\delta\Gamma^{\lambda}_{\sigma \rho}\,$ stands for
the difference between the Christoffel symbols of the Kerr-AdS
spacetime and those of its reference background. The difference
between the metric determinants $\,\delta g=0\,$. Having performed
straightforward evaluation of the above integral with respect to
the Killing vector (\ref{newkill}) and over a $2$-sphere at
infinity we obtain the relation
\begin{eqnarray}
M^{\prime}&=& K [\partial_t -\frac{a}{l^2}\,\partial_{\phi}] =
K\left[\partial_t \right] -
\frac{a}{l^2}\,K\left[\partial_{\phi}\right]
\label{actrel}
\end{eqnarray}
with $$ K\left[\partial_t
\right]=\mathcal{M}^{\prime}\,\,,~~~~~~~~~K\left[\partial_{\phi}\right]=-
{J}^{\prime} \,\,,$$ so that the  mass  of the Kerr-AdS metric
(\ref{4kads}) is given by
\begin{eqnarray}
M^{\prime}&=& \mathcal{M}^{\prime}+ \frac{a}{l^2}\,{J}^{\prime}=
\frac{M}{\Xi^2}\,\,. \label{physmass4}
\end{eqnarray}
It is  important to note that with the angular momentum in
(\ref{4mj}), only this mass  satisfies the first law of
thermodynamics \cite{gpp1}.

\subsection{Electric Charge}

We shall now assume that the  Kerr-AdS black holes described above
may also possess a small test electric charge. The effect of the
electromagnetic field of this charge on the spacetime geometry can
be neglected and the spacetime can still be well described by the
unperturbed metric (\ref{4kads}). In the asymptotically flat case,
the associated solution of the source-free Maxwell equations is
constructed using the well-known fact that for Ricci-flat metrics
a Killing one-form  is closed and co-closed \cite{af,ac}. This
implies that the Killing one-form can be used as a potential
one-form for a test Maxwell field  in the Ricci-flat metrics.
However, the Kerr-AdS metric under consideration is not
Ricci-flat.  Fortunately, one can still use the Killing isometries
to describe the electromagnetic field of a Kerr-AdS black hole.
Namely, it is straightforward to show that {\it the Killing
one-form  $ \delta \hat\xi_{(t)} $ that represents the difference
between the timelike isometries of the Kerr-AdS metric and those
of its reference background can  be used as a potential one-form
for the electromagnetic field.} The Killing one-form $\,\hat
\xi_{(t)} $ is associated with the timelike Killing vector in
(\ref{killing}) and, as before, we use the metric (\ref{4kads})
with $M =0 $ as a reference background. We seek a potential
one-form
\begin{equation}
A= \alpha \, \delta \hat\xi_{(t)}\,\,, \label{potform1}
\end{equation}
where  the constant parameter $\,\alpha\,$   is determined from
examining the Gauss flux
\begin{equation}
Q^{\,\prime}= {\frac{1}{4 \pi}} \oint \,^{\star}F\,\,
\label{gauss}
\end{equation}
which yields
\begin{equation}
\alpha = -\frac{Q}{2 M} \,. \label{fixpara}
\end{equation}
The desired  potential one-form is  thus given by
\begin{equation}
A= -\frac{Q \,r}{\Sigma}\left(dt- \frac{a
\sin^2\theta}{\Xi}\,d\phi \right)\,\,, \label{potform2}
\end{equation}
where the  parameter $ Q $  is related to the electric charge of
the black hole by
\begin{equation}
Q^{\,\prime}=  \frac{Q}{\Xi} \,\,.\label{physcharge}
\end{equation}
The associated  electromagnetic field two-form is given by
\begin{eqnarray}
\label{2form}
 F&=& \frac{Q \left(\Sigma- 2 r^2 \right)
}{\Sigma{^2}} \left(dt-\frac{a \sin^2\theta}{\Xi} d\phi
\right)\wedge dr +
 \frac{Q r a \sin 2\theta}{\Sigma^2} \left(a\, dt - \frac{r^2+a^2}{\Xi}\,d\phi \right)
 \wedge d\theta \,.\nonumber\\
& &
\end{eqnarray}
For some further purposes, it is also useful to calculate the
nonzero contravariant components of the electromagnetic field
tensor. We find that
\begin{eqnarray}
F^{\,01}&=&- \frac{Q\,(r^2+a^2)}{\Sigma\,^3}\,\left(\Sigma- 2 r^2
\right)\,\,, ~~~~~~~~~ F^{\,02}= - \frac{Q a^2 r
\sin2\theta}{\Sigma\,^3}\,\,,
\nonumber \\[3mm]
F^{\,13}& =&  \frac{Q a \left(\Sigma- 2 r^2 \right)}
{\Sigma\,^3}\,\Xi \,, ~~~~~~~~~~~~~~~~~~~~ F^{23} = \frac{2 Q a
r}{\Sigma\,^3}\,\Xi \cot\theta \,\,. \label{emtcontra}
\end{eqnarray}
We recall once again that we have employed here the test-charge
approximation that led to the potential one-form given in
(\ref{potform2}). In the next section we consider the arbitrary
values of the electric charge.

\section{Kerr-Newman-AdS black holes in four dimensions}

Let us now consider a rotating AdS black hole with an arbitrary
electric charge. In this case we need to solve the coupled system
of the Einstein-Maxwell equations to construct the spacetime
metric with associated electromagnetic fields. It is
straightforward to verify that the metric (\ref{4kads}) and the
potential one-form (\ref{potform2}) solve the Einstein-Maxwell
equations
\begin{equation}
{R_{\mu}}^{\nu}= 2\left(F_{\mu \lambda} F^{\nu
\lambda}-\frac{1}{4}\,{\delta_{\mu}}^{\nu}\,F_{\alpha \beta}
F^{\alpha \beta}\right) - 3l^{-2}{\delta_{\mu}}^{\nu}\,\,,
\label{einmax}
\end{equation}
$$\partial_{\nu} (\sqrt{-g}\,F^{\mu \nu})= 0 \,\,, $$
provided that the mass parameter of the metric is re-scaled as
\begin{equation}
M\rightarrow M- \frac{Q^2}{2 r}\,\,.\label{resc}
\end{equation}
Thus, the metric (\ref{4kads}) with the new function
\begin{equation}
\Delta_r = \left(r^2 + a^2\right)\left(1 +\frac{r^2}{l^2}\right) -
2 M r + Q^2\, \label{mdelta}
\end{equation}
goes over into  the familiar Kerr-Newman-anti-de Sitter solution
\cite{carter2} in which  the electromagnetic field is given by the
same potential one-form as (\ref{potform2}). Furthermore, it is
easy to show that all the salient features of the Kerr-AdS metric
given in equations (\ref{angvelocity1}) - (\ref{einvelocity})
remain valid for the Kerr-Newman-AdS solution as well.

Next, we  describe the horizons of the Kerr-Newman-AdS black
holes. The radii of the horizons are determined by the real roots
of the equation
\begin{equation}
\Delta_r = 0 \,.\label{quartic}
\end{equation}
This is a quartic equation which in general has four roots
satisfying Vieta's relations
\begin{eqnarray}
r_1+r_2+r_3+r_4&=&0  \,\,,\nonumber \\
r_1 r_2+r_1 r_3+r_1 r_4+r_2 r_3+r_2 r_4+r_3 r_4&=& a^2+l^2\,\,,\\
r_1 r_2 r_3+ r_1 r_2 r_4+r_2 r_3 r_4+r_1 r_3 r_4&=& 2 M l^2  \,\,,\nonumber \\
r_1 r_2 r_3 r_4&=& l^2 \left( a^2+ Q^2 \right) \,\,.\nonumber
\label{Vieta}
\end{eqnarray}
From these relations we observe that the quartic equation must
allow two real and  a pair of complex conjugate roots. The largest
of the real roots $r_1=r_{+}$  corresponds to the radius of the
black hole's outer event horizon, while the other real root
$r_2=r_{-}$ represents the radius of the inner Cauchy horizon. The
real solutions of equation (\ref{quartic}) can be written in a
compact form using a real root $\,u\,$ of its resolvent cubic
equation. For the real root of the resolvent equation we find
\begin{eqnarray}
u&=&\,\frac{l^2+a^2}{3}+\frac{l^{\,4/3}\,\left(M_{1e}^2-M_{2e}^2
\right)^{2/3}}{\left(2 N^2- M_{1e}^2-M_{2e}^2 \right)^{1/3}} +
l^{\,4/3} \left(2 N^2- M_{1e}^2-M_{2e}^2 \right)^{1/3}\,\,,
\label{resolventr}
\end{eqnarray}
where we have introduced two extreme mass parameters
\begin{eqnarray}
M_{1e}&= &(l/\sqrt{54})\,\sqrt{\zeta + \eta^3} \,\,\,,~~~~~
M_{2e}= (l/\sqrt{54})\, \sqrt{\zeta -\eta^3}\,\,. \label{exmasses}
\end{eqnarray}
Here
\begin{equation}
\label{zeta} \zeta = \left(1+\frac{a^2}{l^2}\right)\left[\frac{36
\left(a^2+Q^2 \right)}{l^2}-\left(1+\frac{a^2}{l^2}\right)^2
\right],~~~ \eta =\left[\left(1+\frac{a^2}{l^2}\right)^2+ \frac{12
\left(a^2+Q^2\right)}{l^2}\right]^{1/2} \label{eta}\,
\end{equation}
and
\begin{equation}
N^2=M^2+\sqrt{\left(M^2-M_{1e}^2\right)\left(M^2-M_{2e}^2\right)}\,\,.
\label{censor1}
\end{equation}
It is easy to show that  only the mass parameter  $ M_{1e} $ has a
definite physical meaning. For vanishing cosmological constant, $
l\rightarrow \infty \,$, one finds that $ M_{1e}^2\rightarrow
a^2+Q^2 $, while $ M_{2e}^2 \rightarrow ~ -\infty \,.$ The
expression for the extreme mass $ M_{1e} $ in (\ref{exmasses})
agrees with that given in Ref. \cite{cck}. It is clear that the
black hole mass parameter $ M $ must satisfy the relation
\begin{equation}
 M\geq M_{1e} \,\,,
 \label{censor2}
\end{equation}
where the equality corresponds to an extreme black hole. The
horizons are located at the radii
\begin{eqnarray}
r_{+}&=& \frac{1}{2}\left(X+Y \right)\,\,,~~~~~~~r_{-}=
\frac{1}{2}\left(X-Y \right)\,\,, \label{horizons}
\end{eqnarray}
where
\begin{eqnarray}
X&=& \sqrt{u-l^2-a^2}\,\,,~~~~~ Y= \sqrt{-u -l^2-a^2 +\frac{4 M
l^2}{X}}\,\,.
\end{eqnarray}
For $\,a=0\, $ and $\,Q=0\,, $ the inner horizon disappears, $\,r
_{-} =0\,$,  and $\,r_{+}\,$ represents the event horizon of a
Schwarzschild-AdS black  hole \cite{stuchlik}. Expanding the above
expressions in powers of $ 1/\,l $  with $ M/\,l\ll 1 $, we find
\begin{eqnarray}
r_{+}&=& \tilde{r}_{+} -\frac{\tilde{r}_{+}^{2}}{2\, l^2}
\,\,\frac{2
M \tilde{r}_{+} - Q^2}{\tilde{r}_{+}-M } + \mathcal{O}\left(\frac{1}{l^4}\right)\,\,,\\[2mm]
r_{-}&=& \tilde{r}_{-} -\frac{\tilde{r}_{-}^{2}}{2\, l^2}
\,\,\frac{2 M \tilde{r}_{-} - Q^2}{\tilde{r}_{-}-M }+
\mathcal{O}\left(\frac{1}{l^4}\right)\,\,, \label{limits1}
\end{eqnarray}
where
$$ \tilde{r}_{\pm}= M\pm \sqrt{M^2-a^2-Q^2}\,\,.$$
We see that the location of the event horizon lies in the range $
r_{-}< r_{+}< \tilde{r}_{+} \,$.

It is also worth emphasizing that for $\,M=M_{1e}\,$ the outer and
inner horizons merge to form an extreme black hole located at the
radius
\begin{eqnarray}
r_{eh}&=&\frac{l}{\sqrt{6}}\,\left(\eta
-1-\frac{a^2}{l^2}\right)^{1/2} . \label{exhorizon}
\end{eqnarray}
In the critical limit of rotation in which $ a^2 = l^2 $ and for
$\,Q = 0\,$, from equations (\ref{exmasses}) and (\ref{exhorizon})
we obtain that
\begin{eqnarray}
\tilde r_{eh}&=& \frac{3}{8}\,M =\frac{l}{\sqrt{3}}\,\,\,.
\label{critical}
\end{eqnarray}
This gives the limiting size for the horizon  of the extreme black
hole. For $\,a^2 < l^2\,$ we find that $\, r_{eh} < \tilde
r_{eh}\,$.  In the critical limit  $ a^2 = l^2 $,  the angular
velocity in (\ref{einvelocity}) becomes
\begin{equation}
\omega_{H}=\frac{1}{l}\,\, \label{crit}
\end{equation}
regardless of the horizon size. As we have mentioned above, this
is the same as the angular velocity of the Einstein universe on
the boundary of AdS spacetime. Thus, the boundary is rotating at
the speed of light $\,(v=\omega \,l \rightarrow 1)\,$ when $\,a^2
\rightarrow l^2\,.$

\subsection{Gyromagnetic Ratio}

An important characteristic of the Kerr-Newman-AdS black hole is
its gyromagnetic ratio. We recall that one of the remarkable facts
about a Kerr-Newman black hole in asymptotically flat spacetime is
that it can be assigned a gyromagnetic ratio $\,g=2\,$\,, just as
an electron in the Dirac theory \cite{carter1}. The parameter
$\,g\,$ is defined as a constant of proportionality in the
equation for the magnetic dipole moment
\begin{equation}
\mu= g\, \frac{Q\,J}{2 \,M}\,\,, \label{g1}
\end{equation}
where $\,M\,$ is the mass, $\,J\,$ is the angular momentum and
$\,Q\,$ is the electric charge of the Kerr-Newman black hole. Here
we wish to calculate the value of the gyromagnetic ratio when the
black hole has  an asymptotic AdS behavior. We  begin with the
associated magnetic dipole moment. The most direct way to
determine it is to examine the asymptotic behavior of the magnetic
field generated by a Kerr-Newman-AdS black hole. For this purpose,
it is useful to introduce an orthonormal tetrad frame which is
given by the basis one-forms
\begin{eqnarray}
\label{basis}
 e^{0} &=&\left(\frac{\Delta_r }{\Sigma
}\right)^{1/2}\left(dt- \frac{a \sin^2\theta}{\Xi }\,d\phi
\right) \,, \nonumber \\[2mm]
e^{3} &=&\left(\frac{\Delta_{\theta} }{\Sigma }\right)^{1/2}\,\sin
\theta \left(a \,dt- \frac{r^2+a^2}{\Xi }\,d\phi
\right)\,\,, \\[2mm]
 e^{1} &=&\left(\frac{\Sigma}{\Delta_r
}\right)^{1/2}dr\,\,,~~~~~e^{2}=
\left(\frac{\Sigma}{\Delta_{\theta }}\right)^{1/2} d\theta  \,.
\nonumber
\end{eqnarray}
The remarkable property of this frame is that an observer at rest
in it measures only the radial components of the electric and
magnetic fields \cite{carter1}. Writing the electromagnetic
two-form (\ref{2form}) in this frame and using the relation
(\ref{physcharge}), we find the following asymptotic expansions
for the radial fields
\begin{eqnarray}
\label{eradials}
 E_{\,\hat r} &=&\frac{Q^{\,\prime}\Xi}{r^2}
+\mathcal{O}\left(\frac{1}{r^4}\right) \,\,,\\[3mm]
 B_{\,\hat r} &=&\frac{2 Q^{\,\prime} a
\Xi}{r^3}\,\cos\theta +\mathcal{O}\left(\frac{1}{r^5}\right)
\,\,.\label{mradials}
\end{eqnarray}
It is easy to check that the  Gaussian flux of the radial electric
field gives the correct value for the electric charge  of the
black hole, see Eq. (\ref{physcharge}). The second equation for
the dominant behavior of the radial magnetic field shows that the
black hole can be assigned a magnetic dipole moment given by
\begin{eqnarray}
\mu ^{\,\prime}&=&Q^{\,\prime}a=\frac{\mu}{\Xi}\,\,,
\label{physmagm}
\end{eqnarray}
where $\,\mu= Q a\,$ is the magnetic dipole moment parameter.
Defining now the $g$-factor in terms of the actual mass, angular
momentum and electric charge, we have  a relation similar to
(\ref{g1}); that is,
\begin{equation}
\mu^{\,\prime}= g\, \frac{ Q^{\,\prime}J^{\,\prime}}{2
M^{\,\prime}}\,\,\,. \label{g2}
\end{equation}
It follows that the Kerr-Newman-AdS  black holes must have a
gyromagnetic ratio corresponding  to $\,g=2\,$ just as the usual
Kerr-Newman black holes in asymptotically flat spacetime. We
recall that for Kerr-Newman black holes in de Sitter space one
must take  $\,l^2 \rightarrow -l^2\,$, which does not affect the
value of the gyromagnetic ratio.

\section{Higher dimensional charged Kerr-AdS black holes}

The higher dimensional generalization of the Kerr-Newman-AdS
solution has not yet been found. However, one can still examine
some electromagnetic properties of rotating AdS black holes in
higher dimensions employing the test-charge approach described in
Sec. II. We  therefore consider a weakly charged black hole and
use the spacetime geometry described by the higher-dimensional
Kerr-AdS solution found in Refs. \cite{hhtr, glpp1}. We first
focus on the Kerr-AdS metric with a single angular momentum in $
N+1 $ dimensions with $ N\geq 3 $ . In the Boyer-Lindquist type
coordinates it is given by
\newpage
\begin{eqnarray}
ds^2 & = & -{{\Delta_r}\over {\Sigma}} \left(\,dt - \frac{a
\sin^2\theta}{\Xi}\,d\phi\,\right)^2 + {\Sigma \over~ \Delta_r}
dr^2 + {\Sigma \over ~\Delta_{\theta}}\,d\theta^{\,2} \nonumber
\\[3mm] &&
+ \,\frac{\Delta_{\theta}\sin^2\theta}{\Sigma} \left(a\, dt -
\frac{r^2+a^2}{\Xi} \,d\phi \right)^2 + r^2 \cos^2{\theta} \,
d\Omega_{N-3}^2\,\,\,. \label{allkads}
\end{eqnarray}
This metric satisfies the Einstein field equations with the
cosmological term, $ R_{\mu \nu}=-N l^{-2}\,g_{\mu \nu} $. The
curvature radius $\, l\,$ of the AdS space is related to the
negative cosmological constant by  $\,\Lambda=- \frac{1}{2}
 N(N-1)\,l^{-2} $,  the metric functions are the same as in
(\ref{metfunct}) except for
\begin{eqnarray}
\Delta_r &= &\left(r^2 + a^2\right)\left(1 +\frac{r^2}{l^2}\right)
- m \,r^{4-N} \,,
\end{eqnarray}
and
\begin{equation}
d\Omega_{N-3}^2 =d {\chi_{1}}^2+ \sin^2{\chi_{1}}\,(\,d
{\chi_{2}}^2+\sin^2{\chi_{2}}\,(...d{\chi_{N-3}}^2...)\,)\,\,,
\label{sphmetric}
\end{equation}
which is the metric on a unit $\,(N-3)\,$-sphere. Here $ m $ is
the mass parameter that reduces to $ 2 M $ for $ N=3 $. For the
determinant of the metric (\ref{allkads}) we find
\begin{equation}
\sqrt{-g}= \frac{\Sigma
\sin\theta}{\Xi}\,\sqrt{\gamma}\,r^{\,N-3}\,
\cos^{\,N-3}{\theta}\,\,,\label{determinant}
\end{equation}
where $\gamma$  stands for the determinant of the metric
(\ref{sphmetric}).  Clearly,  the metric (\ref{allkads}) admits
the same  timelike and  rotational Killing vectors as those given
in (\ref{killing}). In a similar way to (\ref{angvelocity1}), one
can define for the locally nonrotating observers an angular
velocity that at the horizon of the black hole reduces to
\begin{equation}
\Omega_{H}= \frac{a\,\Xi}{r_{+}^2 + a^2}\,\,, \label{hvelocityall}
\end{equation}
where $ r_{+} $ is the radius of the horizon, i.e. the largest
root of equation $ \Delta_r=0 $.

Next, we evaluate the mass and angular momentum of the metric
(\ref{allkads}). Following the four-dimensional case in Sec. II,
we first employ the Komar integrals which in $\,N+1\,$ dimensions
have the form
\begin{eqnarray}
\mathcal{M}^{\prime}&=& - \frac{1}{16\pi\,}\,\frac{N-1}{N-2}\oint
\,^{\star}d(\delta \hat \xi_{(t)})\,\,,~~~~~~~~~~~J^{\prime}=
\frac{1}{16 \pi\,}\oint \,^{\star}d(\delta \hat
\xi_{(\phi)})\,\,.\label{komarall}
\end{eqnarray}
We recall that in performing the above integrals one must again
integrate the differences $ \delta \hat \xi  $ between the Killing
isometries of the metric under consideration and its reference
background. The latter is given by the metric (\ref{allkads}) with
vanishing mass parameter $(m=0)$. Substituting into these
integrals the asymptotic expansions
\begin{eqnarray}
\delta \xi_{(t)}^{t\,;\,r}& = & \frac{m\, (N-2)}{2\,r^{N-1}} +
\mathcal{O}\left(\frac{1}{r^{N+1}}\right)\,\,,
\nonumber \\[4mm]
\delta \xi_{(\phi)}^{t\,;\,r}& = & - \,\frac {a m \, N}{\Xi}\,
\frac{\sin^2\theta}{2\,r^{N-1}}+
\mathcal{O}\left(\frac{1}{r^{N+1}}\right)\,\,. \label{expall}
\end{eqnarray}
and performing  integration over a  $ (N-1)$-sphere at $ r
\rightarrow \infty $   we obtain
\begin{eqnarray}
\mathcal{M}^{\prime}&=& \frac{m (N-1)A_{N-1}}{16\pi\, \Xi}\,
\,\,,~~~~~~ J^{\prime}= \frac{a m
A_{N-1}}{8\pi\,\Xi^2}\,\,,\label{mjall}
\end{eqnarray}
where $ A_{N-1}= 2\,\pi^{\,N/2}/{\,\Gamma(N/2)} $ is the area of a
unit $\,(N-1)\,$-sphere. We note that the obtained  value for the
angular momentum can be regarded as the actual angular momentum
for the Kerr-AdS black holes. This is unambiguously confirmed in
the framework of various methods \cite{hhtr, gpp1, dkt}. However,
with this angular momentum the obtained mass does not satisfy the
first law of thermodynamics. As in the four-dimensional case, the
consistent mass can be found if one uses a new timelike Killing
vector instead of $\,\partial_t\,$ that is nonrotating at $ m=0 $.
We take the new Killing vector in the same form as that given in
(\ref{newkill}) and again employ  the superpotential technique of
Katz, Bi\v c\' ak and Lynden-Bell \cite{kbl}. A straightforward
calculation of the integral (\ref{kblpot}) over a $(N-1)$-sphere
at infinity shows that the relation in (\ref{actrel}) holds in the
higher-dimensional case as well. Thus, for the actual mass  we
have the relation
\begin{eqnarray}
M^{\prime}&=& \mathcal{M}^{\prime}+
\frac{a}{l^2}\,{J}^{\prime}\,\,. \label{physmassall1}
\end{eqnarray}
Finally, we find that the mass and angular momentum of the
Kerr-AdS metric are given by the expressions
\begin{eqnarray}
M^{\prime}&=& \frac{m^{\prime} A_{N-1}}{16\pi\,}\,\left[\,2+
(N-3)\,\Xi\, \right]\,\,,~~~~~~~J^{\prime}= \frac{j^{\,\prime}
A_{N-1}}{8\pi}\,\,, \label{newmjall}
\end{eqnarray}
where we have defined the specific mass and angular momentum as
\begin{eqnarray}
m^{\prime}&=& \frac{m}{\Xi^2}\,\,,~~~~~~~~~~ j^{\,\prime}= \frac{a
m}{\Xi^2}\,\,,\label{specificmj}
\end{eqnarray}
which are reminiscent of the corresponding relations for the mass
and angular momentum  of the Kerr-AdS metric in four dimensions.
We note that the expression  for the mass in (\ref{newmjall})
agrees with that appearing in \cite{gpp1} when adapting the latter
to the case of a single angular momentum.

We turn now to the description of the electromagnetic field
generated by a test electric charge of a higher-dimensional
Kerr-AdS black hole. By the same arguments as those given in Sec.
II, one can construct the vector potential of the electromagnetic
field using the difference between the timelike generators in the
metric (\ref{allkads}) and in its reference background. This gives
rise to the potential one-form
\begin{equation}
A= -\frac{Q\,r^{4-N}}{(N-2)\,\Sigma}\left(dt- \frac{a
\sin^2\theta}{\Xi}\,d\phi \right)\,\,. \label{hpotform}
\end{equation}
The parameter $ Q $ is related to the electric charge of the black
hole by Gauss's law in $ N+1 $ dimensions
\begin{equation}
Q^{\,\prime}= \frac{1}{A_{N-1}} \oint \,^{\star}F\,\,.
\label{gaussall}
\end{equation}
For the electric charge, we find the same relation as that given
in (\ref{physcharge}); that is,
\begin{equation}
Q^{\,\prime}=  \frac{Q}{\Xi} \,\,.\label{Nphyscharge}
\end{equation}

The basis one-forms for the metric (\ref{allkads}) can be chosen
as
\begin{eqnarray}
\label{basisall}
 e^{0} &=&\left(\frac{\Delta_r }{\Sigma
}\right)^{1/2}\left(dt- \frac{a \sin^2\theta}{\Xi }\,d\phi
\right) \,, \nonumber \\[2mm]
e^{1} &=&\left(\frac{\Sigma}{\Delta_r
}\right)^{1/2}dr\,\,,~~~~~e^{2}=
\left(\frac{\Sigma}{\Delta_{\theta }}\right)^{1/2} d\theta
\,\,,\\[2mm]
e^{3} &=&\left(\frac{\Delta_{\theta} }{\Sigma }\right)^{1/2}\,\sin
\theta \left(a \,dt- \frac{r^2+a^2}{\Xi }\,d\phi
\right)\,\,, \nonumber\\[2mm]
e^{4} & =& r \cos\theta\,d\,\chi_1 \,\,,~~~~~~ e^{5}  = r
\cos\theta\,\sin\chi_1 d\,\chi_2 \,\,\nonumber
\end{eqnarray}
and so on. As in the four-dimensional case, the significance of
these basis one-forms is that they define a natural orthonormal
frame in which the electromagnetic field tensor takes its simplest
form. It is straightforward to show that the electromagnetic field
two-form written in terms of basis one-forms (\ref{basisall}) is
given by
\begin{eqnarray}
\label{2formall}
 F&=&- \frac{Q\,r^{3-N}}{(N-2)\,\Sigma{\,^2}} \,\left\{\, \left[(N-2) \,\Sigma -2\, a^2
 \cos^2\theta\right]
 \,e^{0}\wedge e^{1}
 -2 a\, r \cos\theta \,e^{3}\wedge e^{2}
 \right\}\,\,,
\end{eqnarray}
which involves the radial component of the electric field and only
one component of the magnetic field. The dominant behavior of
these fields at spatial infinity is given by
\newpage
\begin{eqnarray}
\label{eradialsall}F_{\hat{r}\hat{t}}& = &\frac{Q^{\,\prime}\,\Xi}
{r^{N-1}}
+\mathcal{O}\left(\frac{1}{r^{N+1}}\right) \,\,,\\[3mm]
F_{\hat{\theta}\hat{\phi}} &= &\frac{2\, Q^{\,\prime} a\,\Xi
}{N-2}\,\,\frac{\cos\theta}{r^N}+\mathcal{O}\left(\frac{1}{r^{N+2}}\right)
\,\,.\label{mradialsall}
\end{eqnarray}
Again, the Gauss flux of the radial electric field  confirms the
value of the electric charge  in (\ref{Nphyscharge}).  We see that
the magnetic field  is determined by the quantity
\begin{eqnarray}
\mu^{\,\prime}&=&\frac{j^{\,\prime}\,Q^{\,\prime}}{m^{\,\prime}}
=\frac{a Q}{\Xi}\,\,,\label{magmom}
\end{eqnarray}
which can be thought of as the magnetic dipole moment of the black
hole. Using the expressions in (\ref{newmjall}) we can rewrite the
magnetic dipole moment in the form
\begin{eqnarray}
\mu^{\,\prime}& = & \left[\,2+ (N-3)\,\Xi\, \right]
\frac{J^{\,\prime}\,Q^{\prime}}{2\,M^{\prime}}\,\,. \label{gyro1}
\end{eqnarray}
Comparing now this expression with that given in (\ref{g2}) we
read off the value of the gyromagnetic ratio
\begin{equation}
g = 2+ (N-3)\,\Xi\, \,\, \label{gyroall}
\end{equation}
for the Kerr-AdS black holes carrying a test electric charge and a
single angular momentum in all higher dimensions. For $\,N=3\,$,
this expression shows that  $\,g=2\,$ is a universal feature of
four dimensions. For vanishing cosmological constant,
$\,l\rightarrow\infty\,$, it recovers the value of the
gyromagnetic ratio found for weakly charged Myers-Perry metrics
\cite{af,aliev1}. However, the most striking feature of the
expression (\ref{gyroall}) appears in the critical limit $\,\,\Xi
\rightarrow 0\,,\,$ in which the boundary Einstein universe
rotates at the speed of light. We see that $\,g\rightarrow 2\,$
irrespective of the spacetime dimension. It is interesting to note
that in a recent work, Ref. \cite{cvetic}, it has been argued that
at the critical limit of rotation the Kerr-AdS black holes are
related to SUSY configurations. That is, a supersymmetric black
hole in an AdS background has the same angular velocity with
respect to a frame that is nonrotating at infinity as that given
in (\ref{crit}). One may therefore conclude that the
supersymmetric black hole must have the gyromagnetic ratio $g=2$ .

\subsection{Alternative calculations}

The value of the gyromagnetic ratio found above can be proved by
alternative calculations using  a different approach. For this
purpose, we define the twist of a Killing one-form  that is
associated with the timelike Killing vector $\,\partial_{t}\,$ of
the metric (\ref{allkads}). This is given by the $\,(N-2)\,$-form
\begin{equation}
{\hat \omega}_{N-2}  =
\frac{1}{(N-2)}\,^{\star}\left({\hat\xi}_{(t)} \wedge
d\,{\hat\xi}_{(t)}\right)\,\,. \label{twist1}
\end{equation}
Physically, this quantity measures the failure of the Killing
vector to be hypersurface orthogonal. For the metric
(\ref{allkads}), we find
\begin{eqnarray}
\label{twist2} {\hat \omega}_{N-2}&=&-\frac{a  m
\cos^{N-3}\theta}{N-2} \left\{ \left[\frac{(N-2)\Sigma- 2 a^2
\cos^2\theta}{\Sigma^2}+\frac{2 r^{N-2}}{m l^2}
\right]\right. \nonumber \\[2mm]  & &  \left.
\sin\theta d\theta + 2\cos\theta \left(\frac {r}{\Sigma^2}-
\frac{r^{N-3}}{m l^2} \right)dr \right\}\wedge d\Sigma_{N-3}\,\,,
\end{eqnarray}
where
\begin{eqnarray}
\label{twistpot1}
d\Sigma_{N-3}=\frac{\sqrt{\gamma}}{(N-3)\,!}\,\,\epsilon_{i_{1}\,i_{2}\,...\,i_{N-3}}\,\,
dx^{i_{1}}\wedge dx^{i_{2}}\wedge...\wedge dx^{i_{N-3}}\,\,.
\end{eqnarray}
It is straightforward to show that the twist form is closed, $\,d
{\hat \omega}_{N-2}=0\,$, implying the existence (locally) of the
twist potential $ (N-3)$-form
\begin{eqnarray}
\label{twistpot2} {\hat \Omega}_{N-3}=\frac{\, a
\cos^{N-2}\theta}{N-2} \left( \frac{m}{\Sigma}
+\frac{2}{l^2}\,\frac{r^{N-2}}{N-2}\right)d\Sigma_{N-3}\,.
\end{eqnarray}
We see that this quantity is not zero for vanishing mass parameter
$m\rightarrow 0 $, reflecting the fact that the associated
background spacetime is indeed rotating at infinity. Performing in
(\ref{twistpot2}) the background subtraction, we obtain the ``
physical " twist
\begin{equation}
\delta \hat\Omega_{N-3}  = \frac{a m}{\Sigma}
\,\,\frac{\cos^{N-2}\theta}{N-2}\,\, d\Sigma_{N-3}\,\,.
\label{twistpot3}
\end{equation}

Next, we  define the magnetic field $\,(N-2)\,$-form
\begin{eqnarray}
{\hat B}_{N-2} &=& i_{\hat\xi_{(t)}}\, ^{\star}F =\,
^{\star}\left({\hat\xi_{(t)}}\wedge F\right)\,\,, \label{mform}
\end{eqnarray}
which in the Kerr-AdS spacetime under consideration has the form
\begin{eqnarray}
{\hat B}_{N-2}&=& \frac{a Q}{\Sigma^2}\,\,\frac{
\cos^{N-3}\theta}{N-2}\,\left\{\left[\,(N-2)\Sigma- 2\,a^2
\cos^2\theta \right] \sin\theta\,d\theta + 2 r \cos\theta\, dr
\right\}\wedge d\Sigma_{N-3}\,\,. \label{mformexp}
\end{eqnarray}
It is easy to verify that one can  also introduce the magnetic
potential $(N-3)$-form  by the equation
\begin{equation}
{\hat B}_{N-2}= -d\,\varphi_{N-3} \,\,, \label{magpot}
\end{equation}
where
\begin{equation}
\varphi_{N-3} = \frac{a Q}{\Sigma}
\,\,\frac{\cos^{N-2}\theta}{N-2}\,\, d\Sigma_{N-3}\,\,.
\label{varphi}
\end{equation}
From a comparison of this expression with (\ref{twistpot3}), it
follows that the asymptotic behavior of the magnetic potential $
(N-3)$-form determines the magnetic moment parameter $ \mu = Q a
$, just as the twist potential $ (N-3) $-form determines the
specific angular momentum parameter $ j=a m  $. This fact can also
be expressed in the form
\begin{equation}
\varphi_{N-3} =  \frac{Q}{m}\,\delta \hat\Omega_{N-3} \,\,,
\label{proofrel}
\end{equation}
which is equivalent to the relation (\ref{magmom}). This proves
the value of the gyromagnetic ratio in (\ref{gyroall}).

\section{General Kerr-AdS black holes in five dimensions}

The general metric for Kerr-AdS black holes with two independent
rotation parameters in five dimensions was first given in
\cite{hhtr}. The simplest form of the metric is given by
\begin{eqnarray}
ds^2 & = & -{{\Delta_r}\over {r^2\,\Sigma}} \left(\,dt - \frac{a
\sin^2\theta}{\Xi_a}\,d\phi\, - \frac{b
\cos^2\theta}{\Xi_b}\,d\psi \right)^2 + \Sigma
\left(\frac{r^2}{\Delta_r}\,dr^2  + \frac{d\theta^{\,2}}{
~\Delta_{\theta}}\right)\, \nonumber
\\[3mm] &&
+ \,\frac{\Delta_{\theta}\sin^2\theta}{\Sigma} \left(a\, dt -
\frac{r^2+a^2}{\Xi_a} \,d\phi \right)^2 +
\frac{\Delta_{\theta}\cos^2\theta}{\Sigma} \left(b\, dt -
\frac{r^2+b^2}{\Xi_b} \,d\psi \right)^2 \nonumber
\\[3mm] &&
+\,\frac{1+r^2\,l^{-2}}{r^2 \Sigma } \left( a\, b \,dt - \frac{b
(r^2+a^2) \sin^2\theta}{\Xi_a}\,d\phi\, - \frac{a (r^2+b^2)
\cos^2\theta}{\Xi_b}\,d\psi \right)^2\,\, ,\label{5dkads}
\end{eqnarray}
where
\begin{eqnarray}
\Delta_r &= &\left(r^2 + a^2\right)\left(r^2 +
b^2\right)\left(1+\frac{r^2}{l^2}\right)
 - m  r^2 \,,~~~~\Sigma = r^2+ a^2 \cos^2\theta + b^2 \sin^2\theta \,\,,
\nonumber \\[2mm]
\Delta_\theta & = & 1 -\frac{a^2}{l^2} \,\cos^2\theta
-\frac{b^2}{l^2} \,\sin^2\theta \,,~~~~~~~~~~~~~~~~~~~ \Xi_a=1 -
\frac{a^2}{l^2}\,\,,~~~~ \Xi_b=1 - \frac{b^2}{l^2}\,\,
\label{5metfunc}
\end{eqnarray}
and  $\,a \,$ and $\,b \,$ are two independent rotation
parameters. The metric determinant is given by
\begin{equation}
\sqrt{-g}= \frac{r \Sigma \sin\theta\,\cos\theta}{\Xi_a\,
\Xi_b}\,\,. \label{5ddeterminant}
\end{equation}
It is also clear that the metric admits  three commuting Killing
vector fields
\begin{equation}
{\xi}_{(t)}= \frac{\partial}{\partial t}\,, ~~~~ {\xi}_{(\phi)}=
\frac{\partial}{\partial \phi} \, , ~~~~ { \xi}_{(\psi)}=
\frac{\partial}{\partial \psi }\,, \label{5dkilling}
\end{equation}
which reflect stationarity and bi-azimuthal symmetry of this
spacetime. The various scalar products of these Killing vectors
are given by
\begin{eqnarray}
{ \xi}_{(t)} \cdot {\xi}_{(t)}&=& g_{tt}= -1+ \frac{m}{\Sigma} -
\frac{r^2 +a^2 \sin^2\theta + b^2 \cos^2 \theta}{l^2}\,\,, \nonumber \\[3mm]
{ \xi}_{(\phi)} \cdot {\xi}_{(\phi)}&= &g_{\phi\phi}= \frac{\sin^2
\theta}{\Xi_{a}^2}\left[\left(r^2+a^2\right) \Xi_a + \frac{m\,a^2
\sin^2 \theta}{\Sigma}\right]\,\,,
\nonumber \\[3mm]
{\xi}_{(\psi)} \cdot {\xi}_{(\psi)}&= &g_{\psi \psi}= \frac{\cos^2
\theta}{\Xi_{b}^2}\left[\left(r^2+b^2\right) \Xi_b + \frac{m\,b^2
\cos^2 \theta}{\Sigma}\right]\,\,,
 \\[3mm]
{\xi}_{(t)} \cdot {\xi}_{(\phi)}&=& g_{t\phi} =  - \frac{a \sin^2
\theta}{\Xi_a}\,\left(\frac{m}{\Sigma}-\frac{r^2+a^2}{l^2}\right)\,\,,
\nonumber \\[3mm]
{\xi}_{(t)} \cdot {\xi}_{(\psi)}&=& g_{t\psi} =  - \frac{b \cos^2
\theta}{\Xi_b}\,\left(\frac{m}{\Sigma}-\frac{r^2+b^2}{l^2}\right)\,\,,
\nonumber \\[3mm]
{\xi}_{(\phi)} \cdot {\xi}_{(\psi)}&=& g_{\phi \psi}= \frac{m
\,a\, b \sin^2 \theta\,\cos^2 \theta\,}{\Sigma \, \Xi_a\, \Xi_b}\
\,\,.\nonumber \label{5dkproduct}
\end{eqnarray}
It proves useful to define a family of locally nonrotating
observers. We recall that in four dimensions a locally nonrotating
observer has a vector of four-velocity that is orthogonal to the
axial Killing vector. Similarly, in the five-dimensional Kerr-AdS
metric we can also define a unit vector of five-velocity for a
locally nonrotating observer. It is given by
\begin{equation}
u^{\mu}=u^{\mu}(r,\theta)=  C \left(
\xi_{(t)}^{\mu}+\Omega_{a}\xi_{(\phi)}^{\mu}
+\Omega_{b}\xi_{(\psi)}^{\mu}\right)\, ,
\end{equation}
where the parameter  $ C $ is fixed  by the normalization
condition $ u^2=-1 $. By analogy, we require that $ u\cdot
{\xi}_{(\phi)}=0 $ and $ u\cdot {\xi}_{(\psi)}=0 $; that is,
\begin{eqnarray}
g_{t \phi}\,u^{t} + g_{\phi\phi}\,u^{\phi} +
g_{\phi\psi}\,u^{\psi} & =& 0\,, \\ [3mm] g_{t \psi}\,u^{t} +
g_{\psi\psi}\,u^{\psi} + g_{\psi\phi}\,u^{\phi} & =& 0 \,\,.
\label{eqs1}
\end{eqnarray}
These equations determine the coordinate angular velocities of the
observers. Straightforward calculations show that they are given
by
\begin{eqnarray}
\label{5dangve2a}
 \Omega_{a} & =& \frac{u^{\phi}}{u^t}\; = \;
\frac{g_{t\psi}\,g_{\phi\psi}-g_{t\phi}\,
g_{\psi\psi}}{g_{\phi\phi}\,g_{\psi\psi}-{g_{\phi\psi}}^2}\,
=\,\frac{a \,\Xi_a \left[m\left(r^2+b^2 \right) \Delta_{\theta}
-l^{-2}\,\Delta_r \Sigma \,\Xi_b \right]}{m
\left(r^2+a^2\right)\left(r^2+b^2 \right) \Delta_{\theta} +
\Delta_r \Sigma \,\Xi_a \,\Xi_b}
 \,\,, \\ [5mm]
\Omega_{b} & =& \frac{u^{\psi}}{u^t}\; =
\;\frac{g_{t\phi}\,g_{\phi\psi}
-g_{t\psi}\,g_{\phi\phi}}{g_{\phi\phi}\,g_{\psi\psi}-{g_{\phi\psi}}^2}\,
= \frac{b \,\Xi_b \left[m\left(r^2+a^2 \right) \Delta_{\theta}
-l^{-2}\,\Delta_r \Sigma \,\Xi_a \right]}{m
\left(r^2+a^2\right)\left(r^2+b^2 \right) \Delta_{\theta} +
\Delta_r \Sigma \,\Xi_a \,\Xi_b} \,\,. \label{5dangve2b}
\end{eqnarray}
For vanishing cosmological constant, $ l \rightarrow \infty \,$,
these expressions reduce to those obtained in \cite{af}. Far from
the black hole we have
\begin{eqnarray}
\Omega_{a} & =& -\frac{a}{l^2} + \frac{a
m}{r^4}\,\frac{\Delta_{\theta}}{\Xi_a \Xi_b} +
\,\mathcal{O}\left(\frac{1}{r^6}\right)\,\,, ~~~~~ \Omega_{b} =
-\frac{b}{l^2} + \frac{b\,m}{r^4}\,\frac{\Delta_{\theta}}{\Xi_a
\Xi_b} + \,\mathcal{O}\left(\frac{1}{r^6}\right) \,\,.
\label{5dangvel3}
\end{eqnarray}
We see that the angular velocities of the locally nonrotating
observers in both $\,\phi\,$ and $\,\psi\,$ $2$-planes of rotation
do not vanish at spatial infinity; the {\it bi-dragging of
inertial frames} occurs at spatial infinity as well. When
approaching the black hole horizon, $ \Delta_r \rightarrow 0 $, it
follows  from equations (\ref{5dangve2a}) and (\ref{5dangve2b})
that
\begin{equation}
\Omega_{a(+)} = \frac{a\,\Xi_a}{r_{+}^2 +a^2}\,\,,~~~~~~~
\Omega_{b(+)} = \frac{b\,\Xi_b}{r_{+}^2 +b^2}\,\,,
\label{5dhorizons}
\end{equation}
where $\, r_{+}\,$ is the radius of the outer event horizon. These
quantities can be interpreted as angular velocities of the event
horizon in two independent orthogonal 2-planes of rotation
\cite{hhtr}. One can also verify that the Killing vector
\begin{equation}
\chi = {\bf \xi}_{(t)}+ \Omega_{a(+)}\,{\bf \xi}_{(\phi)}+
\Omega_{b(+)} \,{\bf \xi}_{(\psi)}\,\,,\label{5dkhorizon}
\end{equation}
which is a linear combination of the Killing vectors in
(\ref{5dkilling}), correctly describes the isometry properties of
the horizon geometry. It becomes tangent to the null surface of
the event horizon, thereby showing that the Killing horizon is the
same as the event horizon of the five-dimensional Kerr-AdS metric.

The Komar mass and angular momenta of the metric (\ref{5dkads})
are obtained  by employing the integrals (\ref{komarall}) in the
five-dimensional case. We find the expressions
\begin{eqnarray}
\mathcal{M}^{\prime}&=& \frac{3 \pi m}{8\,\Xi_a \Xi_b}\,
\,\,,~~~~~~ J^{\,\prime}_a= \frac{\pi a m} {4 \,\Xi_{a}^2\,
\Xi_b}\,\,,~~~~~~~J^{\,\prime}_b= \frac{\pi b m} {4\,\Xi_{a}\,
\Xi_{b}^2}\,\,. \label{5mjj}
\end{eqnarray}
These are easily verified by using the asymptotic expansions
\begin{eqnarray}
\delta \xi_{(t)}^{t\,;\,r}& = & \frac{m}{r^3} +
\mathcal{O}\left(\frac{1}{r^{5}}\right)\,\,,
\nonumber \\[4mm]
\delta \xi_{(\phi)}^{t\,;\,r}& = & - \,\frac {2 a m }{\Xi_a}\,
\frac{\sin^2\theta}{r^{3}}+
\mathcal{O}\left(\frac{1}{r^{5}}\right)\,\,,
\nonumber \\[4mm]
\delta \xi_{(\psi)}^{t\,;\,r}& = & - \,\frac {2 b m }{\Xi_b}\,
\frac{\cos^2\theta}{r^{3}}+
\mathcal{O}\left(\frac{1}{r^{5}}\right)\,\, \label{5dexpall}
\end{eqnarray}
and performing integration over a 3-sphere at spatial infinity. As
in the previous cases, the Komar expressions for the angular
momenta are unambiguous.  Again, with these angular momenta the
expression for the mass does not agree with the first law of
thermodynamics \cite{gpp1}. The agreement is achieved for a mass
that is associated with the timelike Killing vector
\begin{equation}
\partial_t -\frac{a}{l^2}\,\,\partial_{\phi}-\frac{b}{l^2}\,\,\partial_{\psi} \,\,,\label{5dnonrkill}
\end{equation}
which has a vanishing twist in the $ m=0 $ reference background.
Evaluating the integrals in (\ref{kblpot}) with respect to this
Killing vector, we obtain the expression
\begin{eqnarray}
M^{\prime}&=& K \left[\partial_t
-\frac{a}{l^2}\,\partial_{\phi}-\frac{b}{l^2}\,\partial_{\psi}\right]
= K\left[\partial_t \right] -
\frac{a}{l^2}\,K\left[\partial_{\phi}\right]
-\frac{b}{l^2}\,K\left[\partial_{\psi}\right]
\,\,,\label{5dactrel}
\end{eqnarray}
where $$ K\left[\partial_t
\right]=\mathcal{M}^{\prime}\,\,,~~~~~~~~~K\left[\partial_{\phi}\right]=-
J^{\,\prime}_a\,\,,~~~~~~~~~ K\left[\partial_{\psi}\right]=-
J^{\,\prime}_b \,\,.$$ Using  the expressions  in (\ref{5mjj}), we
obtain the desired actual mass of the five dimensional Kerr-AdS
metric
\begin{equation}
M^{\prime}=  \frac{\pi m \left(2 \,\Xi_a +2\, \Xi_b-
\Xi_a\,\Xi_b\right)}{8\,\Xi_{a}^2\,\Xi_{b}^2}
\,\,.\label{5dphysmass}
\end{equation}
This formula is in agreement with that obtained in \cite{gpp1}
from purely thermodynamic considerations.

We shall now describe the electromagnetic field that is generated
by a test electric charge of the Kerr-AdS black hole. Again, the
corresponding vector potential can be constructed using the
timelike isometries of the metric (\ref{allkads}) and those of its
$ m=0 $ reference background.  Using the expressions
(\ref{potform1}) and (\ref{gaussall}) in the case under
consideration, we find that  the potential one-form is given by
\begin{equation}
A= -\frac{Q}{2\,\Sigma}\,\left(dt- \frac{a
\sin^2\theta}{\Xi_a}\,d\phi -\frac{b \cos^2\theta}{\Xi_b}\,d\psi
\right)\,\,, \label{5dpotform1}
\end{equation}
where  $ Q $ is determined by  the electric charge of the black
hole through the relation
\begin{equation}
Q^{\,\prime}=\frac{Q}{\Xi_a\,\Xi_b} \,\,.\label{5dcharge}
\end{equation}
For the electromagnetic field two-form we find
\begin{eqnarray}
\label{5d2form}
 F&=&- \frac{Q\, r}{\Sigma^2}\left(dt- \frac{a
\sin^2\theta}{\Xi_a}\,d\phi -\frac{b \cos^2\theta}{\Xi_b}\,d\psi
\right)\wedge dr +
 \frac{Q a \sin 2\theta}{2 \Sigma^2} \left(a \,dt - \frac{r^2+a^2}{\Xi_a}\,d\phi \right)
 \wedge d\theta \nonumber\\[2mm]
& & -\,\frac{Q b \sin 2\theta}{2 \Sigma^2} \left(b \,dt -
\frac{r^2+b^2}{\Xi_b}\,d\psi \right)
 \wedge d\theta\,\,.
\end{eqnarray}
For some purposes, it is also useful to know the nonzero
contravariant components of the electromagnetic field tensor. They
are given by
\begin{eqnarray}
F^{tr}&=&\frac{Q\,(r^2+a^2)(r^2+b^2)}{r\,\Sigma\,^3}\,\,,
~~~~~~~~~~~ F^{t\theta}=
-\frac{Q\,(a^2-b^2)\,\sin2\theta}{2\,\Sigma\,^3}\,\,,
\nonumber \\[3mm]
F^{r\phi}& =& - \frac{Q\,a\,\Xi_a\,(r^2+b^2)}{r\,\Sigma\,^3}\,\,,
~~~~~~~~~~~~~~ F^{r\psi} = - \frac{Q\,b\,\Xi_b
(r^2+a^2)}{r\,\Sigma\,^3}\,\,,
 \nonumber \\[3mm]
F^{\theta\phi}& =&\frac{Q\,a\,\Xi_a \cot\theta}{\Sigma\,^3}\,\,,
~~~~~~~~~~~~~~~~~~~~~ F^{\theta\psi} = - \frac{Q\,b\,\Xi_b
\tan\theta}{\Sigma\,^3}\,\,. \label{5demtcontra}
\end{eqnarray}
It is clear that a five-dimensional charged  Kerr-AdS black hole
must have two independent magnetic dipole moments due to its
rotations in two independent orthogonal $2$-planes. To determine
the value of these magnetic dipole moments, we first need to
generalize  the Carter frame given in (\ref{basis}) to include the
case of the five-dimensional metric (\ref{5dkads}). Rather lengthy
calculations show that the desired frame is given by the basis
one-forms
\begin{eqnarray} \label{g5dkadsbasis}
e^{0} &=& \frac{1}{r} \left(\frac{\Delta_r}{\Sigma
}\right)^{1/2}\left(dt- \frac{a \sin^2\theta}{\Xi_a}\,d\phi -
\frac{b \cos^2\theta}{\Xi_b}\,d\psi \right) \,,
\nonumber \\[2mm]
e^{1} &=&r \left(\frac{\Sigma}{\Delta_r
}\right)^{1/2}dr\,\,,~~~~~e^{2}=
\left(\frac{\Sigma}{\Delta_{\theta}}\right)^{1/2}\, d\theta
\,\,,\nonumber \\[2mm]
e^{3} &=&\frac{\sin \theta }{r \Sigma^{1/2}}\left\{\left[B- b^2
\cos^2\theta \left(1+ \frac{r^2}{l^2}\right) \frac{A}{Z}\right]
\left(a\, dt- \frac{r^2+a^2}{\Xi_a}\,d\phi\right)
\right. \nonumber \\[2mm]  & &  \left.
+\, a b \cos^2\theta \left(1+\frac{r^2}{l^2}\right)\frac{A}{Z}
\,\left( b\, dt- \frac{r^2+b^2}{\Xi_b}\,d\psi\right)
 \right\}\,\,,  \\[2mm]
e^{4} &=&\frac{\cos \theta }{r \Sigma^{1/2}}\left\{\left[A- a^2
\sin^2\theta \left(1+ \frac{r^2}{l^2}\right) \frac{B}{Z}\right]
\left(b \,dt- \frac{r^2+b^2}{\Xi_b}\,d\psi\right)
\right. \nonumber \\[2mm]  & &  \left.
+\, a b \sin^2\theta \left(1+\frac{r^2}{l^2}\right)\frac{B}{Z}
\,\left( a\, dt- \frac{r^2+a^2}{\Xi_a}\,d\phi\right)
 \right\}\,\,, \nonumber
\end{eqnarray}
where
\begin{eqnarray}
A&=&\left[\Delta_{\theta} r^2 + a^2 \left(1+
\frac{r^2}{l^2}\right)\right]^{1/2},~~~ B=\left[\Delta_{\theta}
r^2 + b^2 \left(1+ \frac{r^2}{l^2}\right)\right]^{1/2},\nonumber \\[2mm]
Z&= & r \Sigma^{1/2}{\Delta_{\theta}}^{1/2} + A B\,\,.
\end{eqnarray}
In this frame, the electromagnetic field two-form (\ref{5d2form})
becomes
\begin{eqnarray}
\label{2form5dkads} F&=&- \frac{Q\,r}{\Sigma{\,^2}}\,e^{0}\wedge
e^{1} + \frac{Q a \cos\theta}{\Sigma{\,^2}}\,f\, e^{3}\wedge e^{2}
 -\frac{Q b \sin\theta}{\Sigma{\,^2}}\,h\, e^{4}\wedge e^{2}\,\,,
\end{eqnarray}
where
\begin{eqnarray}
f&=&\frac{r {\Delta_{\theta}}^{1/2} A + \Sigma^{1/2}
B}{Z}\,\,,~~~~~~~~ h= \frac{r {\Delta_{\theta}}^{1/2} B +
\Sigma^{1/2} A}{Z}\,\,.
\end{eqnarray}
In the asymptotic region, $ r \rightarrow \infty $, and for $ a=0
$ or $ b=0 $ as well as  in the special case $ a=b $, the
functions $ f $ and $ h $  tend to unity, while for arbitrary
values of the rotation parameters they  depend on the angle
$\theta$ and the dimensionless ratios of the rotation parameters
to the curvature radius of the  AdS background. Thus, from the
asymptotic behavior of  the electromagnetic field two-form in
(\ref{2form5dkads}), it follows that the Kerr-AdS black hole can
be assigned  two distinct magnetic dipole moments
\begin{eqnarray}
\mu^{\,\prime}_{(a)} & = & \frac{Q a}{\Xi_a\,\Xi_b} \,,~~~~~~
\mu_{(b)}^{\,\prime}  =  \frac{Q b}{\Xi_a\,\Xi_b} \,,
\label{dipole}
\end{eqnarray}
where we have used relation (\ref{5dcharge}). For vanishing
cosmological constant these expressions are in agreement with
those obtained in \cite{af,aliev1}. Using expressions in
(\ref{5mjj}) and  (\ref{5dphysmass}), it is easy to see that the
magnetic dipole moments are expressed in terms of the mass and
angular momenta of the black hole as follows
\begin{eqnarray}
\mu_{(a)}^{\,\prime}& = & \left(2 - \Xi_a + 2
\,\frac{\Xi_a}{\Xi_b} \right) \frac{J^{\,\prime}_a\,Q^{\prime}}{2
M^{\prime}}\,\,\,,~~~~~~~\mu_{(b)}^{\,\prime} = \left(2 - \Xi_b +
2 \,\frac{\Xi_b}{\Xi_a} \right)
\frac{J^{\,\prime}_b\,Q^{\prime}}{2 M^{\prime}}\,\,.\label{5gyro1}
\end{eqnarray}
Comparing  these expressions with  the definition of the
gyromagnetic ratio in (\ref{g2}), we find the two distinct
gyromagnetic ratios
\begin{eqnarray}
g_{(a)}& = & 2 - \Xi_a + 2 \,\frac{\Xi_a}{\Xi_b}\,\,\,,~~~~~~~~
g_{(b)} =  2 - \Xi_b + 2 \,\frac{\Xi_b}{\Xi_a}\,\,. \label{gyros5}
\end{eqnarray}
in accordance with two independent rotations of the Kerr-AdS black
hole in five dimensions. First of all, we note that for zero
cosmological constant, $ \Xi_a \rightarrow1 $ and  $ \Xi_b
\rightarrow 1 $, the two gyromagnetic ratios merge into one value
$ g=3 $ \cite{af}. For a single angular momentum, $ a=0 $ or $ b=0
$, we have  the gyromagnetic ratio corresponding to that given by
(\ref{gyroall}). In the special case with two equal angular
momenta, the gyromagnetic ratios again merge leading to the value
\begin{equation}
g =  4 - \Xi \,\,.
\label{equalgyros5}
\end{equation}
It is interesting to note that in the critical limit
$\Xi\rightarrow 0 $, where the Einstein universe rotates at the
speed of light, the gyromagnetic ratio tends  $g= 4$, in contrast
to the case of a single angular momentum  for which
$\,g\rightarrow 2\,$ (see Sec. IV).

\section{Conclusion}

In this paper we have discussed the electromagnetic properties of
rotating charged black holes in anti-de Sitter spacetime in four
and higher dimensions focusing basically on test electromagnetic
fields. The test field approach has enabled us to employ an
elegant way of constructing the associated solutions of the
source-free Maxwell field equations. Namely, we have shown that
the difference between the timelike generators in the Kerr-AdS
spacetime and in the corresponding reference background can be
taken as a potential one-form for the electromagnetic field. In
four dimensions, the test potential one-form is the same as that
describing the full electromagnetic field of the familiar
Kerr-Newman-AdS  black hole. We have also solved the quartic
equation determining the positions of horizons in the
Kerr-Newman-AdS metric and found analytic formulas for the radii
of the horizons. The gyromagnetic ratio for these black holes
turns out to be $\,g=2\,$, the same value as for asymptotically
flat black holes.

Turning to the case of Kerr-AdS black holes carrying a single
angular momentum and a test electric charge in all higher
dimensions, we have re-derived the expressions for the mass and
angular momentum that are consistent with the first law of
thermodynamics. Exploring then the asymptotic behavior of the
electromagnetic fields of these black holes we have determined
their gyromagnetic ratio. The use of thermodynamically consistent
expressions for the mass and angular momentum  has led to the
value of the gyromagnetic ratio that crucially depends on the
dimensionless ratio of the rotation parameter to the curvature
radius of the AdS  background. The striking feature of this
dependence appears for maximally rotating black holes for which
the gyromagnetic ratio approaches  $ g=2 $  regardless of the
spacetime dimension. In this case the boundary of the AdS
spacetime is rotating at the speed of light.

Finally, we have examined a general five-dimensional Kerr-AdS
black hole involving two independent angular momenta. We have
given the precise expressions for the angular velocities of
locally nonrotating observers in the spacetime and re-derived its
thermodynamically consistent mass and angular momenta. Moreover,
we have derived the potential one-form that describes the
electromagnetic field generated by a test electric charge of the
black hole. For the five-dimensional Kerr-AdS metric we have also
defined a natural orthonormal frame in which the electromagnetic
field two-form takes its simplest form. This is a generalization
of the familiar Carter frame in four-dimensional Kerr-Newman
metric. As expected, a five-dimensional charged Kerr-AdS black
hole possesses two distinct magnetic dipole moments due to its
rotation in two orthogonal 2-planes. We have shown that the black
hole has two distinct gyromagnetic ratios, unlike its
asymptotically flat counterpart for which the $g$-factor is always
equal to $3$. The gyromagnetic ratios merge  in the special case
of two equal angular momenta. Furthermore, in this case the
gyromagnetic ratio tends to $ g=4 $ for the maximum values of the
angular velocities, in contrast to the single angular momentum
case where $ g =2 $ .

\section{Acknowledgments}

The author thanks the Scientific and Technological Research
Council of Turkey (T{\"U}B\.{I}TAK) for partial support under the
Research Project 105T437. He also thanks \\G. W. Gibbons and B.
Mashhoon for useful comments.

\end{document}